\documentclass[]{spie}  

\usepackage{cite}
\makeatletter
\let\NAT@parse\undefined
\makeatother
\usepackage{amsmath,amssymb,amsfonts,mathtools}
\usepackage{algorithmic}
\usepackage{graphicx}
\usepackage{textcomp}
\usepackage{caption}
\usepackage{float}
\usepackage{array}
\usepackage{tabularx}
\usepackage{textcomp}
\usepackage{siunitx}
\usepackage{hyperref}
\hypersetup{
colorlinks=true,
linkcolor=blue,
citecolor=blue,
filecolor=magenta,      
urlcolor=cyan,
pdftitle={},
pdfpagemode=FullScreen,
}

\title{No-gold-standard evaluation of quantitative imaging methods in the presence of correlated noise}

\author[a]{Ziping Liu}
\author[a]{Zekun Li}
\author[b]{Joyce C. Mhlanga}
\author[b]{Barry A. Siegel}
\author[a,b]{Abhinav K. Jha}
\affil[a]{Department of Biomedical Engineering, Washington University, St. Louis, MO, USA}
\affil[b]{Mallinckrodt Institute of Radiology, Washington University School of Medicine, St. Louis, MO, USA}

\authorinfo{Corresponding author: Abhinav K. Jha (a.jha@wustl.edu)}

\begin{document} 

\begin{titlepage}
This manuscript has been accepted to SPIE Medical Imaging, February 20-24, 2022. Please use the following reference when citing the manuscript.

Liu, Z., Li, Z., Mhlanga, J. C., Siegel, B. A., and Jha, A. K., ``No-gold-standard evaluation of quantitative imaging methods in the presence of correlated noise", Proc. SPIE Medical Imaging, 2022.

\end{titlepage}

\clearpage

\maketitle

\begin{abstract}

Objective evaluation of quantitative imaging (QI) methods with patient data is highly desirable, but is hindered by the lack or unreliability of an available gold standard. 
To address this issue, techniques that can evaluate QI methods without access to a gold standard are being actively developed.
These techniques assume that the true and measured values are linearly related by a slope, bias, and Gaussian-distributed noise term, where the noise between measurements made by different methods is independent of each other. 
However, this noise arises in the process of measuring the same quantitative value, and thus can be correlated.
To address this limitation, we propose a no-gold-standard evaluation (NGSE) technique that models this correlated noise by a multi-variate Gaussian distribution parameterized by a covariance matrix.
We derive a maximum-likelihood-based approach to estimate the parameters that describe the relationship between the true and measured values, without any knowledge of the true values.
We then use the estimated slopes and diagonal elements of the covariance matrix to compute the noise-to-slope ratio (NSR) to rank the QI methods on the basis of precision.
The proposed NGSE technique was evaluated with multiple numerical experiments.
Our results showed that the technique reliably estimated the NSR values and yielded accurate rankings of the considered methods for $83\%$ of $160$ trials.
In particular, the technique correctly identified the most precise method for $\sim 97\%$ of the trials.
Overall, this study demonstrates the efficacy of the NGSE technique to accurately rank different QI methods when the correlated noise is present, and without access to any knowledge of the ground truth.
The results motivate further validation of this technique with realistic simulation studies and patient data.

\end{abstract}

\keywords{no-gold-standard, objective evaluation, medical imaging, quantitative imaging}

\section{INTRODUCTION}
\label{sec:introduction}
Medical imaging provides a mechanism to study \textit{in vivo} physiological properties of the human body, and thus plays an important role in the diagnosis, prognosis, and assessment of treatment response of different diseases. 
To facilitate decision-making in clinical practice, quantitative imaging (QI), i.e., the extraction of numerical or statistical features from medical images, is being actively investigated \cite{sullivan2015metrology,rosenkrantz2015clinical}.
QI has demonstrated substantial promise in multiple clinical applications. 
These include the quantification of metabolic tumor volume from oncological positron emission tomography (PET) for predicting clinical outcomes \cite{ohri2015pretreatment}, quantification of dopamine transporter uptake from single-photon emission computed tomography (SPECT) to assess the severity of Parkinson disease \cite{filippi2005123i,moon2020physics}, and quantification of regional uptake from PET and SPECT for dosimetry in targeted radionuclide therapy \cite{flux2006impact,ljungberg20023,dewaraja2012mird,li2021projection}. 

Given the significant interest in QI, multiple methods have been and are being developed for QI. 
For clinical translation of QI, it is essential that the measurements made by those methods are reliable.
Thus, there is an important need for objective evaluation of the reliability of measurements obtained using QI methods. 
Typically, such evaluation requires the presence of either the true value of the quantitative parameter or a reference standard. 
Such true values or reference standards can be available in realistic simulation and physical phantom studies \cite{du2005partial,jin2013evaluation,ouyang2006fast,he2005monte,liu2021bayesian}. 
While these studies are important for the initial development of QI methods, there is an important need for techniques that can perform objective evaluation of QI methods directly with patient data.
Such evaluation then requires the presence of gold-standard quantitative values.
These are typically time-consuming, expensive, and tedious to obtain.
Further, even when an approximate gold standard is available, it could suffer from the lack of reliability.
Thus, techniques that can objectively evaluate QI methods in the absence of a gold standard are much needed.

To objectively evaluate QI methods without the knowledge of a gold standard, a regression-without-truth (RWT) technique was proposed in a set of seminal papers \cite{hoppin2002objective,kupinski2002estimation}.
The RWT technique assumes that the true and measured values are linearly related by a slope, bias, and Gaussian-distributed noise term.
It was demonstrated that even in the absence of a gold standard, the values of the slope, bias, and the standard deviation of the noise term for all the considered QI methods can be estimated using a maximum-likelihood (ML) approach.
These estimated parameters can then be used to rank different QI methods on the basis of precision.
The efficacy of the RWT technique was demonstrated in evaluating segmentation methods on the task of estimating the apparent diffusion coefficient from diffusion-weighted magnetic resonance imaging (MRI) scans \cite{jha2010evaluating,jha2012task}, and on the task of estimating the left ventricular ejection fraction from cardiac cine MRI sequences \cite{lebenberg2012nonsupervised}.
The RWT technique was then advanced further and the efficacy of the resultant no-gold-standard evaluation (NGSE) technique \cite{jha2016no} was demonstrated in objectively evaluating reconstruction methods for SPECT on the task of quantifying regional uptake \cite{jha2015objective,jha2016no}.
Further, the technique was applied to clinical oncological PET images to evaluate segmentation methods on the task of measuring metabolic tumor volume \cite{jha2017practical}. While the findings from these studies are encouraging, an important assumption in these existing evaluation techniques is that the noise between measurements obtained using different QI methods is independent of each other.
The noise with different QI methods arises in the process of measuring the same true value, and then can be correlated. 
Thus, this assumption is often violated. 
To address this issue, we propose an advanced NGSE technique that accounts for the presence of such correlated noise.
We start by presenting the theory of this technique.

\section{METHODS}

\subsection{Theory}
\label{sec:methods(theory)}
Consider a clinical scenario where a total of $P$ patients are being scanned by an imaging system.
From the acquired data of each patient, a set of $K$ QI methods are used to measure certain quantitative values.
Such quantitative values can be the mean activity concentration within different organs of interest.  
Our objective is to estimate the parameters that can describe the relationship between the true and measured values, without access to a gold standard. 
These estimated parameters can then be used to rank the QI methods.

We assume that there exists a linear stochastic relationship between the true and measured values. 
This relationship is parameterized by a slope, bias, and Gaussian-distributed noise term. 
We note that this noise arises in the process of measuring the same true values but with different methods.
Thus, the noise is expected to be correlated.
We model this correlated noise by a zero-mean multi-variate Gaussian distribution denoted by $\mathcal{N}(0, \mathbf{C})$.
Specifically, the diagonal elements of $\mathbf{C}$, i.e., $\{ \sigma_k^2 \}$, denote the variance of the noise of each method. 
The off-diagonal elements of $\mathbf{C}$, i.e., $\{ \sigma_{k,k'} \}$, denote the covariance of the noise between methods $k$ and $k'$.
For the $p^\mathrm{th}$ patient, denote the true value by $a_p$ and the estimated value using the $k^\mathrm{th}$ method by $\hat{a}_{p,k}$.
Additionally, denote the slope and bias of the $k^\mathrm{th}$ method by $u_k$ and $v_k$, respectively.
For the $p^\mathrm{th}$ patient, we can then write the relationship between the true and measured values as
\begin{align}
    \left[ 
        \begin{array}{c}
             \hat{a}_{p,1} \\ \hat{a}_{p,2} \\ \vdots \\ \hat{a}_{p,K}
        \end{array}
    \right]
    =
    \begin{bmatrix}
        u_1 & v_1 \\ u_2 & v_2 \\ \vdots & \vdots \\ u_K & v_K
    \end{bmatrix}
    \left[
        \begin{array}{c}
             a_p \\ 1
        \end{array}
    \right]
    +
    \mathcal{N}(0, \mathbf{C}).
\label{eq: rel. btw. true and meas. in matrix notation}
\end{align}

Denote the vector $\left[ \hat{a}_{p,1}, \hat{a}_{p,2}, \dots, \hat{a}_{p,K} \right]^T$ by $\boldsymbol{\hat{A}}_p$, the matrix containing $\{u_k\}$ and $\{v_k\}$ by $\boldsymbol{\Theta}$, and the vector $\left[a_p, 1\right]^T$ by $\boldsymbol{A}_p$.
Based on Eq.~\eqref{eq: rel. btw. true and meas. in matrix notation}, obtaining the probability of observing $\boldsymbol{\hat{A}}_p$ given the knowledge of $\{\boldsymbol{A}_p, \boldsymbol{\Theta}, \mathbf{C}\}$ depends on the true values, which are unknown.
To address this issue, we next assume that the true values are sampled from a four-parameter beta distribution (FPBD) parameterized by a vector $\boldsymbol{\mathit{\Omega}}$ \cite{jha2016no}. This FPBD incorporates the fact that the true values lie within a certain range.
Additionally, the FPBD provides the capability to model a wide variety of the ranges and shapes of the true distribution.

Let $\boldsymbol{\mathcal{\hat{A}}} = \{\boldsymbol{\hat{A}}_p, \ p = 1, 2, \dots, P\}$ denote the collection of measurements made by all the $K$ methods from all the $P$ patients.
The NGSE technique uses an ML approach to estimate the values of $\{\boldsymbol{\Theta}, \mathbf{C}, \boldsymbol{\mathit{\Omega}}\}$ that maximize the probability of observing $\boldsymbol{\mathcal{\hat{A}}}$.
The ML estimate of $\{\boldsymbol{\Theta}, \mathbf{C}, \boldsymbol{\mathit{\Omega}}\}$ is given by
\begin{align}
    \left\{\hat{\boldsymbol{\Theta}},\hat{\mathbf{C}},\hat{\boldsymbol{\mathit{\Omega}}}\right\}_{\mathrm{ML}} = 
    \underset{\boldsymbol{\Theta}, \mathbf{C}, \boldsymbol{\mathit{\Omega}}}{\mathrm{arg~max}}
    \left\{ 
    \mathrm{pr}(\boldsymbol{\mathcal{\hat{A}}} | \boldsymbol{\Theta}, \mathbf{C}, \boldsymbol{\mathit{\Omega}}) 
    \right\},
\label{eq: ML estimates}
\end{align}
where $\mathrm{pr}(\boldsymbol{\mathcal{\hat{A}}} | \boldsymbol{\Theta}, \mathbf{C}, \boldsymbol{\mathit{\Omega}})$ denotes the probability of observing $\boldsymbol{\mathcal{\hat{A}}}$ given the knowledge of $\{\boldsymbol{\Theta}, \mathbf{C}, \boldsymbol{\mathit{\Omega}}\}$.
We note that obtaining this probability does not require any knowledge of the true values.

The expression for $\mathrm{pr}(\boldsymbol{\mathcal{\hat{A}}} | \boldsymbol{\Theta}, \mathbf{C}, \boldsymbol{\mathit{\Omega}})$ was determined to obtain the ML estimates using a constrained optimization technique based on the interior-point algorithm \cite{byrd1999interior}.
From the estimated parameters, we used the slope terms $\{\hat{u}_k\}$ and noise standard deviation terms $\{\hat{\sigma}_k\}$, i.e., the square root of the diagonal elements of the covariance matrix, to compute the noise-to-slope ratio (NSR) for each method.
For the $k^\mathrm{th}$ method, the NSR is given by
\begin{align}
    \text{NSR}_k = 
\frac{\hat{\sigma}_k}{\hat{u}_k}.
\end{align}
The NSR evaluates QI methods on the basis of precision \cite{kupinski2002estimation,hoppin2002objective,jha2016no}, and a lower value indicates a more precise estimation performance.

\subsection{Evaluating the NGSE technique using numerical experiments}
\label{sec:methods(evaluation)}
We evaluated the performance of the NGSE technique using multiple numerical experiments. In each experiment, $P = 200$ true values were sampled from a known FPBD. 
From these true values, noisy measured values were generated for $K = 3$ hypothetical QI methods.
Each method yielded outputs that were linearly related to the true values by a slope of $u_k$ and bias of $v_k$.
The variance of the noise of each method was characterized by the diagonal elements of the covariance matrix $\mathbf{C}$.
Additionally, the covariance of the noise between different methods were characterized by the off-diagonal elements of $\mathbf{C}$.
These noisy measurements were then input to the NGSE technique to estimate $\{\boldsymbol{\Theta}, \mathbf{C}, \boldsymbol{\mathit{\Omega}}\}$. 
From these estimated parameters, we used the slope terms $\{\hat{u}_k\}$ and the noise standard deviation terms $\{\hat{\sigma}_k\}$ to compute the NSR for all methods to rank them based on precision, as described in Sec.~\ref{sec:methods(theory)}.

In this evaluation, we sampled the $200$ true values from FPBD for $4$ combinations of $\boldsymbol{\mathit{\Omega}}$ such that different ranges and shapes of the true distribution were modeled. 
To evaluate the sensitivity of the NGSE technique to correlated noise, we generated two sets of QI methods for each combination of $\boldsymbol{\mathit{\Omega}}$.
The first set of methods had lower correlated noise with $\{\sigma_{1,2}, \sigma_{1,3}, \sigma_{2,3}\} = \{0.004,0.008,0.012\}$.
In contrast, the second set of methods had higher correlated noise with $\{\sigma_{1,2}, \sigma_{1,3}, \sigma_{2,3}\} = \{0.015,0.02,0.03\}$.
For both sets, the values of slope $\{u_k\}$, bias $\{v_k\}$, and variance of the noise $\{\sigma_k^2\}$ of the three methods were set to $\{1.1,0.9,1.05\}$, $\{0.1,0.2,0.3\}$, and $\{0.04, 0.09, 0.2025\}$, respectively.
Finally, for each combination of $\{\boldsymbol{\Theta}, \mathbf{C}, \boldsymbol{\mathit{\Omega}}\}$, we repeated the experiment for $20$ different noise realizations. 
Thus, we evaluated the performance of the NGSE technique for a total of $4\times2\times20=160$ trials. 

\section{Results}

We first present the performance of the NGSE technique for the set of hypothetical QI methods that had lower correlated noise.
The means and standard deviations of the estimated slope $\{\hat{u}_k\}$, noise standard deviation $\{\hat{\sigma}_k\}$, and resultant NSR for all considered methods are reported in Table~\ref{tab: Results (numerical): low correlated noise}.
As described in Sec.~\ref{sec:methods(evaluation)}, these statistics were computed from a total of $80$ trials. 
In each trial, we considered either a different combination of $\{\boldsymbol{\Theta}, \mathbf{C}, \boldsymbol{\mathit{\Omega}}\}$ or a different noise realization of the synthetic measurements given the true values.
We observe that the NGSE technique reliably estimated the slope, noise standard deviation, and consequently the NSR values.
From the estimated NSR values, the NGSE technique accurately ranked the methods for $78\%$ of the $80$ trials.
Further, the technique correctly identified method $1$ as the most precise method for $97\%$ of the $80$ trials.

\begin{table*}[h]
\centering
\captionsetup{justification=centering}
\caption{The means and standard deviations of slope, noise standard deviation, and resultant NSR estimated using the NGSE technique for the set of methods that had lower correlated noise.}
\resizebox{\textwidth}{!}{
\begin{tabular}{|c|c|c|c|c|c|c|}
\hline
Method index & True slope & Estimated slope & True noise standard deviation & Estimated noise standard deviation  & True NSR & Estimated NSR \\ \hline
1 & 1.10 & 1.11~$\pm$~0.08 & 0.20 & 0.17~$\pm$~0.09 & 0.18 & 0.16~$\pm$~0.08\\ \hline
2 & 0.90 & 0.89~$\pm$~0.07 & 0.30 & 0.30~$\pm$~0.06 & 0.33 & 0.35~$\pm$~0.09\\ \hline
3 & 1.05 & 1.06~$\pm$~0.09 & 0.45 & 0.44~$\pm$~0.04 & 0.43 & 0.42~$\pm$~0.07\\ \hline
\end{tabular}
}
\label{tab: Results (numerical): low correlated noise}
\end{table*}

We then present in Table~\ref{tab: Results (numerical): high correlated noise} the results for the set of methods that had higher correlated noise.
We again observe that the NGSE technique reliably estimated the slope, noise standard deviation, and resultant NSR.
For $87\%$ of the $80$ trials, the technique yielded accurate rankings of the considered methods. 
Further, the technique correctly identified that method $1$ was the most precise for $97\%$ of the $1,600$ trials.

\begin{table*}[h]
\centering
\captionsetup{justification=centering}
\caption{The means and standard deviations of slope, noise standard deviation, and resultant NSR estimated using the NGSE technique for the set of methods that had higher correlated noise.}
\resizebox{\textwidth}{!}{
\begin{tabular}{|c|c|c|c|c|c|c|}
\hline
Method index & True slope & Estimated slope & True noise standard deviation & Estimated noise standard deviation  & True NSR & Estimated NSR \\ \hline
1 & 1.10 & 1.13~$\pm$~0.09 & 0.20 & 0.17~$\pm$~0.08 & 0.18 & 0.15~$\pm$~0.08\\ \hline
2 & 0.90 & 0.91~$\pm$~0.07 & 0.30 & 0.30~$\pm$~0.06 & 0.33 & 0.34~$\pm$~0.08\\ \hline
3 & 1.05 & 1.07~$\pm$~0.09 & 0.45 & 0.44~$\pm$~0.05 & 0.43 & 0.42~$\pm$~0.07\\ \hline
\end{tabular}
}
\label{tab: Results (numerical): high correlated noise}
\end{table*}

\section{Discussion and Conclusion}

For clinical translation of QI, there is an important need for techniques that can objectively evaluate QI methods with patient data. 
In this context, existing statistical techniques assume that the noise between measurements obtained using different methods is independent of each other.
However, this assumption can often be violated since the noise arises in the process of measuring the same true value. 
To address this issue, we developed an NGSE technique that models this correlated noise by a multi-variate Gaussian distribution.

Our results from the numerical experiments (Tables~\ref{tab: Results (numerical): low correlated noise} and \ref{tab: Results (numerical): high correlated noise}) showed that the NGSE technique yielded reliable estimates of slope, noise standard deviation, and consequently the NSR for all hypothetical QI methods. 
Additionally, the technique yielded accurate rankings of these methods for $83\%$ of the total $160$ trials.
Further, the technique was able to identify the most precise method for $97\%$ of the cases.
This observation is especially important since when evaluating different QI methods, the objective is typically to find the method that yields the most reliable performance \cite{jha2017practical}.
All these results demonstrated that in controlled settings, where the true and measured values were linearly related by design, the NGSE technique was able to reliably rank QI methods without access to any knowledge of the ground truth.

The results motivate further evaluation of the proposed technique with QI methods that are developed for clinical applications.
These include methods developed for reconstruction, post-reconstruction processing, segmentation, and quantification. 
Often, such methods are evaluated using strategies that rely on the availability of a ground truth.
Further, this ground truth may not be relevant to the clinical task. 
For example, segmentation methods are evaluated using metrics such as the Dice score and Hausdorff distance, which quantify spatial overlap and shape similarity, respectively, between the estimated segmentation and a certain ground-truth segmentation. 
Such evaluation then requires access to the true segmentation, which is typically unavailable. 
Usually manual segmentations are used as a surrogate for the ground truth, but these can be erroneous and suffer from intra- and inter-reader variability \cite{leung2020physics}. Similarly, denoising methods for low-dose images are evaluated by comparing the denoised image to a certain normal-dose image using metrics of structural similarity index and root mean square error. However, the normal-dose image is also noisy, and thus provides a limited measure of ground truth. 
More importantly, it is unclear whether the evaluation based on those conventional metrics correlates with the clinical task \cite{yu2020ai,zhu2021comparing,liu2022need}.
Thus, these methods should preferably be evaluated based on clinical-task performance \cite{jha2021objective}.
The proposed NGSE technique provides a mechanism to perform evaluation on clinically relevant quantitative tasks and without access to the ground truth.  

One limitation of the proposed technique is that the true and measured values are assumed to be linearly related. 
This linear relationship is desirable in QI since it ensures that the measured quantitative value is linearly related to the biological effect.
However, this assumption of linearity may not always hold true in QI. 
To address this issue, one strategy is to check whether the measurements made by different methods are linearly related to each other. 
This will increase the confidence that the measured values are also linearly related to the ground truth \cite{jha2017practical}.
A second limitation is that the NGSE technique requires many patient images since multiple parameters need to be estimated. 
One way to reduce the required number of input images is to incorporate the prior information of the parameters to be estimated \cite{jha2015incorporating}.
Thus, extending the proposed technique to incorporate such prior knowledge is an important research direction.

In conclusion, our study demonstrated the ability of the proposed NGSE technique to accurately rank different QI methods in the presence of correlated noise, and without the need for any knowledge of the ground truth. 
The results motivate further evaluation of the technique with realistic simulation studies and patient data.

\section*{ACKNOWLEDGEMENTS}
Financial support for this work was provided by the National Institute of Biomedical Imaging and Bioengineering R01-EB031051, R56-EB028287, and R21-EB024647 (Trailblazer Award). 

\bibliography{report} 
\bibliographystyle{spiebib} 

\end{document}